\documentclass[11pt,preprint]{aastex}

\slugcomment{Accepted for the publication in the {\it Astrophysical Journal Letters}}

\def\farcm{\hbox{$.\mkern-4mu^\prime$}}

\def\la{\mathrel{\hbox{\rlap{\hbox{\lower4pt\hbox{$\sim$}}}\hbox{$<$}}}}
\def\ga{\mathrel{\hbox{\rlap{\hbox{\lower4pt\hbox{$\sim$}}}\hbox{$>$}}}}

\shortauthors{Park}
\shorttitle{G292}

\begin{document}

\title{A Half-Megasecond Chandra Observation of the Oxygen-Rich
Supernova Remnant G292.0+1.8}

\author{Sangwook Park\altaffilmark{1}, John P. Hughes\altaffilmark{2}
Patrick O. Slane\altaffilmark{3}, David N. Burrows\altaffilmark{1},
B. M. Gaensler\altaffilmark{4}, and Parviz Ghavamian\altaffilmark{5}
}

\altaffiltext{1}{Department of Astronomy and Astrophysics, Pennsylvania State
University, 525 Davey Laboratory, University Park, PA. 16802;
park@astro.psu.edu}
\altaffiltext{2}{Department of Physics and Astronomy, Rutgers University,
136 Frelinghuysen Road, Piscataway, NJ 08854-8019; also Department of
Astrophysical Sciences, Princeton University, Peyton Hall,  Princeton, NJ
08544-1001 }
\altaffiltext{3}{Harvard-Smithsonian Center for Astrophysics, 60 Garden Street,
Cambridge, MA. 02138}
\altaffiltext{4}{School of Physics, The University of Sydney, NSW 2006,
Australia}
\altaffiltext{5}{Department of Physics and Astronomy, Johns Hopkins University,
3400 North Charles Street, Baltimore, MD 21218}


\begin{abstract}

We report on our initial analysis of a deep 510 ks observation of the 
Galactic oxygen-rich supernova remnant (SNR) G292.0+1.8 with the {\it Chandra 
X-ray Observatory}. Our new {\it Chandra} ACIS-I observation has a larger field of 
view and an order of magnitude deeper exposure than the previous {\it Chandra} 
observation, which allows us to cover the entire SNR and to detect new metal-rich
ejecta features. We find a highly non-uniform distribution of thermodynamic 
conditions of the X-ray emitting hot gas that correlates well with the optical 
[O {\small III}] emission, suggesting the possibility that the originating 
supernova explosion of G292.0+1.8 was itself asymmetric. We also reveal spectacular
substructures of a torus, a jet, and an extended central compact nebula all 
associated with the embedded pulsar J1124$-$5916.

\end{abstract}


\section {\label {sec:intro} INTRODUCTION}

G292.0+1.8 is one of only three known ``oxygen-rich'' supernova remnants (SNRs)
in the Galaxy \citep{goss79,murdin79}, the others  being Cassiopeia A and Puppis A. 
Optical emission from these SNRs contains fast-moving ($v$ $\ga$ 1000 km s$^{-1}$) 
O-rich ejecta knots, which is generally taken as evidence for He-burning nucleosynthesis 
in the core of massive stars ($>$ 10 M$_{\odot}$) (e.g., Blair et al.\ 2000). O-rich SNRs
provide a rare opportunity for the study of core-collapse supernova (SN) nucleosynthesis 
and the subsequent evolution of the remnant, including in particular the interaction 
of ejecta fragments and the blast wave with circumstellar medium (CSM) produced by
massive stellar winds from the progenitor. G292.0+1.8 is especially intriguing
because it exhibits all the characteristics of a textbook-example core-collapse SNR:
a central pulsar and pulsar wind nebula (PWN), metal-rich ejecta, shocked CSM, and 
the blast wave.

To facilitate a study of SNR G292.0+1.8 in unprecedented detail, we performed
a deep AO7 {\it Chandra} observation during 2006 September 13$-$October 20, with
the SNR centered on the I-array of the Advanced CCD Imaging Spectrometer (ACIS).
Our ACIS-I observation takes advantage of a larger (17$'$$\times$17$'$) field of view 
(FOV) and an order of magnitude deeper exposure (510 ks) than the previous
{\it Chandra} ACIS-S3 observation (8$'$$\times$8$'$ FOV and 43 ks exposure; Hughes et al.
2001). This new deep {\it Chandra} observation is an essential part of our
multi-wavelength campaign on G292.0+1.8, including new observations in the radio, 
infrared, optical, and X-ray bands. Here we report on first results from the new 
{\it Chandra} images: a large-scale overview of the SNR in \S~2 and new 
insights into the nature of the PWN in \S~3. A description of the observations, data
reduction, and results from extensive spectral and imaging analyses of detailed
substructures of the SNR and PWN will be presented elsewhere.

\section{\label{sec:snr} Supernova Remnant}

\begin{figure}[]
\figurenum{1}
\centerline{\includegraphics[angle=0,width=0.8\textwidth]{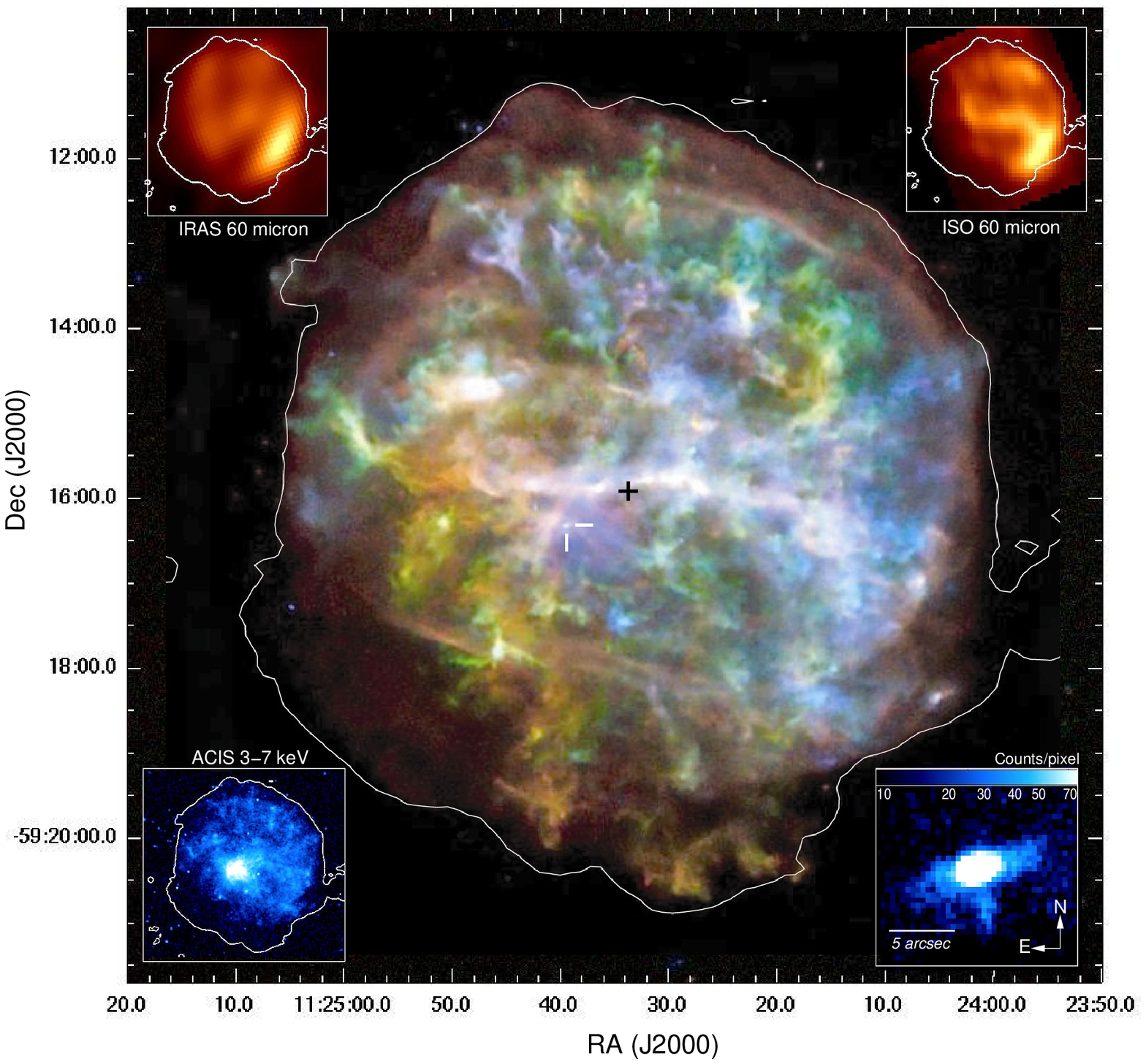}}
\figcaption[]{\small An X-ray-color image of G292.0+1.8. Red is the sum of the 0.58$-$0.71 and 
0.88$-$0.95 keV bands (dominated by emission from O Ly$\alpha$ and Ne He$\alpha$), 
orange is the 0.98$-$1.10 keV band (Ne Ly$\alpha$), green is the 1.28$-$1.43 keV 
band (Mg He$\alpha$), and blue is the sum of the 1.81$-$2.05 and 2.40$-$2.62 keV bands
(Si He$\alpha$ + S He$\alpha$). Each subband image has been exposure-corrected, binned 
with 2$\times$2 pixels ($\sim$1$^{\prime\prime}$), and smoothed for the purposes of 
display. The radio SNR center \citep{gaen03} is marked with a black cross. The position 
of PSR J1124$-$5916 \citep{hughes03} is marked with white bars near the SNR center. 
{\it Lower left} inset is the 3$-$7 keV band image (binned by 4$\times$4 pixel and 
smoothed by a Gaussian with $\sigma$ = 2 pixels). The brightest central parts of the 
PWN is saturated to white. {\it Lower right} inset is the 2$-$7 keV band image of 
the PWN (20$^{\prime\prime}$ $\times$ 20$^{\prime\prime}$ FOV centered on 
the pulsar position). The image is unbinned (0$\farcs$492) and smoothed by a Gaussian 
with $\sigma$ = 1 pixel to emphasize the faint torus and jet. {\it Upper left} and 
{\it upper right} insets are the 60 $\mu$m images from {\it IRAS} Galaxy Atlas 
\citep{cao97} and archival {\it ISO} data, respectively. In all panels except for the 
{\it lower right}, a boundary contour from the 20 cm radio image taken by Australian 
Telescope Compact Array is overlaid. 
\label{fig:fig1}}
\end{figure}

Figure~\ref{fig:fig1} shows an X-ray color image of G292.0+1.8 created from our
deep {\it Chandra} data. The outer boundary of the radio SNR is overlaid with a
white contour. The energy bands displayed in each color were chosen to
emphasize major atomic line emission that illustrates the distribution of electron 
temperatures and ionization states across the SNR (Table~\ref{tbl:tab1}). We note 
that line and underlying continuum emission is included in each subband; e.g.,
the bright blue emission near the SNR's center is dominated by synchrotron
emission from the PWN (see the inset in Figure~\ref{fig:fig1}; Hughes et al.\ 2001).
Our deep ACIS-I exposure comprehensively images the entire SNR to its faint
outermost edge, which matches well the extent of the radio remnant (Figure~\ref{fig:fig1})
and traces the location of the blast wave. Ejecta knots (Park et al.\ 2004, Gonzalez 
\& Safi-Harb 2003) appear brightly colored -- yellow, green or blue -- in 
Figure~\ref{fig:fig1}.  They extend closest to the rim in the west and south and
are furthest away in the SE quadrant. Due to their red-orange color in Figure~\ref{fig:fig1} 
and positional coincidence with [O {\small III}] ejecta \citep{wink06}, the southernmost 
X-ray knots are likely to be O-rich as well. The SNR's full diameter is $\sim$9$\farcm$6 
(N-S) and $\sim$8$\farcm$4 (E-W), corresponding to physical sizes of $\sim$14.7$-$16.7 
$d_6$ pc, where $d_6$ is the distance to G292.0+1.8 in units of 6 kpc, the distance 
we assume throughout \citep{gaen03}.

Our new image of G292.0+1.8 shows little evidence for the featureless, spectrally-hard, 
and geometrically-thin X-ray filaments that trace sites of efficient particle 
acceleration in other young SNRs, such as Cas A, Tycho, and SN 1006.  
Instead the SNR's rim is defined by spectrally-soft emission that is faint and 
diffuse in places (e.g., the SE rim) and filamentary elsewhere. Particle acceleration
is evidently occurring under rather different conditions in G292.0+1.8 than other 
young Galactic SNRs.

The SNR interior contains a complex network of knots and filaments with a variety 
of colors and morphologies.  The overall color distribution of these features is 
highly asymmetric; red-orange emission is dominant in the S-SE, while the W-NW 
regions are bright in emission appearing green-blue in color. We are confident that 
these variations largely reflect differences in the underlying distributions
of gas temperature and ionization in the metal-rich ejecta, rather than variable
foreground absorption.  This is supported by the relative spatial distributions 
of the Ne He$\alpha$ and Ly$\alpha$ lines.  Because these lines are separated in 
energy by only $\sim$0.1 keV, their flux ratio is insensitive to absorption 
variations. The equivalent width maps show that Ne He$\alpha$ emission is 
relatively enhanced in the S-SE, while Ne Ly$\alpha$ emission is enhanced in 
the W-NW \citep{park02}. The metal abundance variation is also unlikely 
a significant contributor for the observed large-scale color variation based on 
our hardness ratio (HR) analysis (see below).

We constructed a HR map (Figure~\ref{fig:fig2}a) to investigate further the variation 
in the thermodynamic state of the ejecta across G292.0+1.8.  
(For technical reasons we added the continuum component [$E$ $<$ 1.25 keV] to the 
soft band line emission to enhance the hard band lines and suppress the strong 
hard continuum emission from the PWN.)  There are prominent enhancements in
Figure~\ref{fig:fig2}a near the W and NW boundary of the SNR that generally trace 
``fingers'' of hot ejecta protruding out to the very boundary of the SNR (green-blue 
filaments in Figure~\ref{fig:fig1}). There is also an overall large-scale variation 
from HR values $\sim$2.4 in the hard ``NW'' region to $\sim$1.1 in the soft
``SE'' region (Figure~\ref{fig:fig2}a).  Example spectra extracted from small 
regions within the larger ``NW'' and ``SE'' regions clearly show the distinctive 
line ratios responsible for the observed HR variation (Figure~\ref{fig:fig2}b). 
We can draw on our previous work fitting the spectra of individual knots to 
relate HR values to plasma temperatures.  Spectral region 3 from Park et al. (2004) 
lies in the hard ``NW'' region, and spectral analysis yields a best-fit $kT \sim 5$
keV, corresponding to HR $\sim$ 2.5. We find that this HR value strongly traces
the electron temperature rather than on individual elemental abundances. 
Although we did not study knots in the soft ``SE'' region previously, our preliminary 
spectral modeling of the SE regions favors significantly lower temperatures 
($kT$ $\sim$ 0.7 keV), corresponding to HR values $\sim$ 1.0. Thus, the observed 
HR distribution reveals a large-scale spatial variation of thermal condition of 
metal-rich ejecta, which could not be detected by previous low-resolution data
\citep{hughes94}; i.e., a significantly higher temperature of the hot gas ($kT$ 
$\sim$ 5 keV) is implied in the N-W regions, while a relatively lower-temperature 
plasma ($kT$ $\la$ 1 keV) prevails in the S-E regions of the SNR.

There is also a very similar and highly significant gradient in the optical 
properties of G292.0+1.8.  The region with the lowest HR values is coincident 
with the bulk of the optical [O {\small III}] emission indicated by 
``SE'' region in Figure~\ref{fig:fig2}a \citep{ghav05,wink06}. Across the projected 
middle of the SNR (within an $\sim$3$'$ wide region aligned roughly N-S),
there are dozens of isolated optical knots (generally uncorrelated with X-ray 
knots), while on the western edge no optical emission appears at all. 
This high degree of correlation between the optical and X-ray properties suggests 
the possibility that radiative cooling in the ejecta is responsible for the SE-NW 
gradient in observed properties. In this picture, the SE ejecta would be undergoing
significant, perhaps catastrophic cooling. Across the projected middle of the
SNR, cooling is probably just beginning to occur in isolated knots that happen to have 
the appropriate thermodynamic conditions, while the emission toward the NW remains 
fully nonradiative.  Variation in the ambient density surrounding G292.0+1.8 
could provide an explanation for the observed asymmetry in the ejecta properties.  
However, there is no evidence for a higher density in the SE regions \citep{braun86} 
as would be expected. In fact, 60 $\mu$m images\footnote{The apparent differences in 
the detailed structure of the two IR images in Figure~\ref{fig:fig1} are unlikely 
to be real because of the highly processed nature of both images. The {\it ISO} image 
was constructed from a rastor scan of many pointings using a $3\times 3$ pixel 
($\sim$45$^{\prime\prime}$ per pixel) IR array that, unfortunately, did not fully 
cover the SE rim of the SNR. The {\it IRAS} image, which did cover the entire SNR, 
was the result of an image reconstruction process intended to improve the angular 
resolution from the native value of several arcminutes to the level of 
1$^{\prime}$$-$2$^{\prime}$. Given these significant differences, the overall general 
agreement between the images, specifically, the brightness of the SW rim and the 
faintness toward the SE, seems reasonable to us.} (see upper right and upper left 
insets to Figure~\ref{fig:fig1}) show that around the rim of the SNR the SE region 
is a minimum in flux, while the SW is a maximum. Thus, albeit somewhat speculative 
at the current stage of the analysis, we raise the possibility that the ejecta 
asymmetry has its origin in some intrinsic asymmetry of the SN explosion itself, 
such as a variation in the density or velocity distribution from SE to NW. Further 
work including detailed X-ray spectral analysis of ejecta and hydrodynamic studies 
appropriate for G292.0+1.8 are required to test this asymmetric explosion scenario.

\section{\label{sec:pwn} Pulsar Wind Nebula}

Previous {\it Chandra} observations of G292.0+1.8 \citep{hughes01} revealed 
a point source, now known to be the pulsar J1124$-$5916 \citep{camilo02,hughes03},
powering an extended synchrotron nebula \citep{torii98} (see the lower left inset 
to Figure~\ref{fig:fig1}). The emission from the pulsar itself was observed to be
surrounded by a compact elliptical structure roughly 1$\farcs$8 $\times$ 
1$^{\prime\prime}$ in size. Our deep ACIS-I observation confirms this 
structure, and reveals additional faint emission suggestive of a jet/torus 
structure ($\sim 5^{\prime\prime}$ in jet length and torus radius, repectively; see 
the lower right inset to Figure~\ref{fig:fig1}) similar to those observed 
to form just outside the pulsar wind termination shock region in a number 
of other young PWNe (e.g., Gaensler \& Slane 2006 and references therein). 
Our preliminary spectral analysis indicates that these 
features show a power law spectrum with photon index $\Gamma$ $\sim$ 1.5$-$1.8, 
supporting the idea that they are synchrotron emission associated with the 
PWN. The ratio between the N-S and E-W sizes of the torus implies 
an inclination of the axis of the torus (with  respect to the line of sight) 
of $\sim$20$^{\circ}$. Physical sizes are then $\sim$0.4 $d_6$ pc for the jet and 
a radius of $\sim$0.15 $d_6$ pc for the torus. This is similar to the size of the 
jet/torus structure observed in 3C~58 \cite{slane02}, although we note that 
there is large variation in the size and relative brightness of such 
structures from system to system.

Since the E-W ``belt'' of the SNR \citep{tuohy82} appears to be a relic feature 
of the progenitor star's equatorial winds \cite{park02}, the pulsar spin-axis
defined by the N-S orientation of the jet is generally aligned with
the progenitor's rotation axis, at least in projection. If the pulsar
position offset to the SE of the radio center of the SNR represents the
direction of the pulsar velocity, a misalignment of $\theta$ $\sim$
70$^{\circ}$ between the projected velocity and spin-axis vectors is implied.

This is in contrast to the strong tendency toward spin-kick alignment claimed 
by Ng \& Romani (2007), although it is important to note that many of their estimated 
velocity vectors are, like ours, based on pulsar offsets from the geometric centers 
of their host SNRs -- a technique that is quite sensitive to the ambient density 
since the SNRs will expand more rapidly in the direction of lower density, thus
creating a shell that is not centered on the explosion point. Without a proper motion 
measurement, our results for J1124$-$5916 are thus not conclusive, although the density 
distribution inferred from the 60 $\mu$m data, which is enhanced in the NW-W-SW 
direction, is difficult to reconcile with the current pulsar position if the observed 
spin axis is aligned with the velocity vector. Finally, we note that our results 
are even consistent with possible orthogonality between the spin and kick directions, 
as might be suggested for some pulsars based on polarization measurements 
\citep{johns05,rankin07}.

\section{\label{sec:con} Summary}

Our deep {\it Chandra} observation of SNR G292.0+1.8 detects the entire outer 
boundary of the blast wave and reveals metal-rich ejecta knots reaching the shock front. 
We find no evidence for strong particle acceleration sites in X-rays. The initial 
large-scale analysis reveals a highly non-uniform distribution of X-ray emitting hot 
gas. This overall structure is caused by an asymmetric distribution of the gas
temperatures, rather than variable foreground absorption. The expected properties 
of the ambient medium required to explain this asymmetry are inconsistent with 
observations, which leads us to the supposition that the supernova explosion itself 
was asymmetric.  A detailed X-ray spectral analysis of the metal-rich ejecta and 
hydrodynamical simulations will be required to test this scenario. Proper motion 
measurements of the blast wave using the new radio data and detailed IR/optical 
studies of thermal states of cooling ejecta will also be important tests.

We discover a torus and a jet associated with the PWN of the embedded pulsar 
J1124$-$5916, similar to a growing number of such structures observed in other young 
pulsar-driven systems.  Assuming that the pulsar's birthplace was at the geometric
center of the radio SNR and that the jet defines the direction of the pulsar spin axis, 
we find a large misalignment of $\sim70^\circ$ between the spin and kick velocity 
directions, in apparent contrast to suggestions of spin-kick alignments in other systems.

\acknowledgments

The authors thank the {\it Chandra} Education and Public Outreach group for
their help to generate the X-ray color image. This work was supported in part
by Smithsonian Astrophysical Observatory under {\it Chandra} grant GO6-7049A
(Penn State), GO6-7049B (Johns Hopkins), and GO6-7049C (Rutgers). P.G. was also
supported by {\it Hubble Space Telescope} grant HST-GO-10916.06-A.

\clearpage

\begin{deluxetable}{cccccc}
\footnotesize
\tablecaption{Energy Bands Used to Generate the X-ray Color
and Hardness Ratio Images.
\label{tbl:tab1}}
\tablewidth{0pt}
\tablehead{ \colhead{Energy Band} & \colhead{Identification} & \colhead{Color} &
\colhead{Energy Band} & \colhead{Identification} & \colhead{Color} \\
 \colhead{(keV)} & & & \colhead{(keV)} & &}
\startdata
0.40$-$0.50 & Continuum & - & 1.16$-$1.25 & Continuum & - \\ 
0.58$-$0.71 & O Ly$\alpha$ & Red & 1.28$-$1.43 & Mg He$\alpha$ & Green \\
0.75$-$0.84 & Continuum & - & 1.81$-$2.05 & Si He$\alpha$ & Blue \\
0.88$-$0.95 & Ne He$\alpha$ & Red & 2.40$-$2.62 & S He$\alpha$ & Blue \\
0.98$-$1.10 & Ne Ly$\alpha$ & Orange & - & - & - 
\enddata

\end{deluxetable}

\clearpage

\begin{figure}[]
\figurenum{2}
\centerline{\includegraphics[angle=0,width=0.8\textwidth]{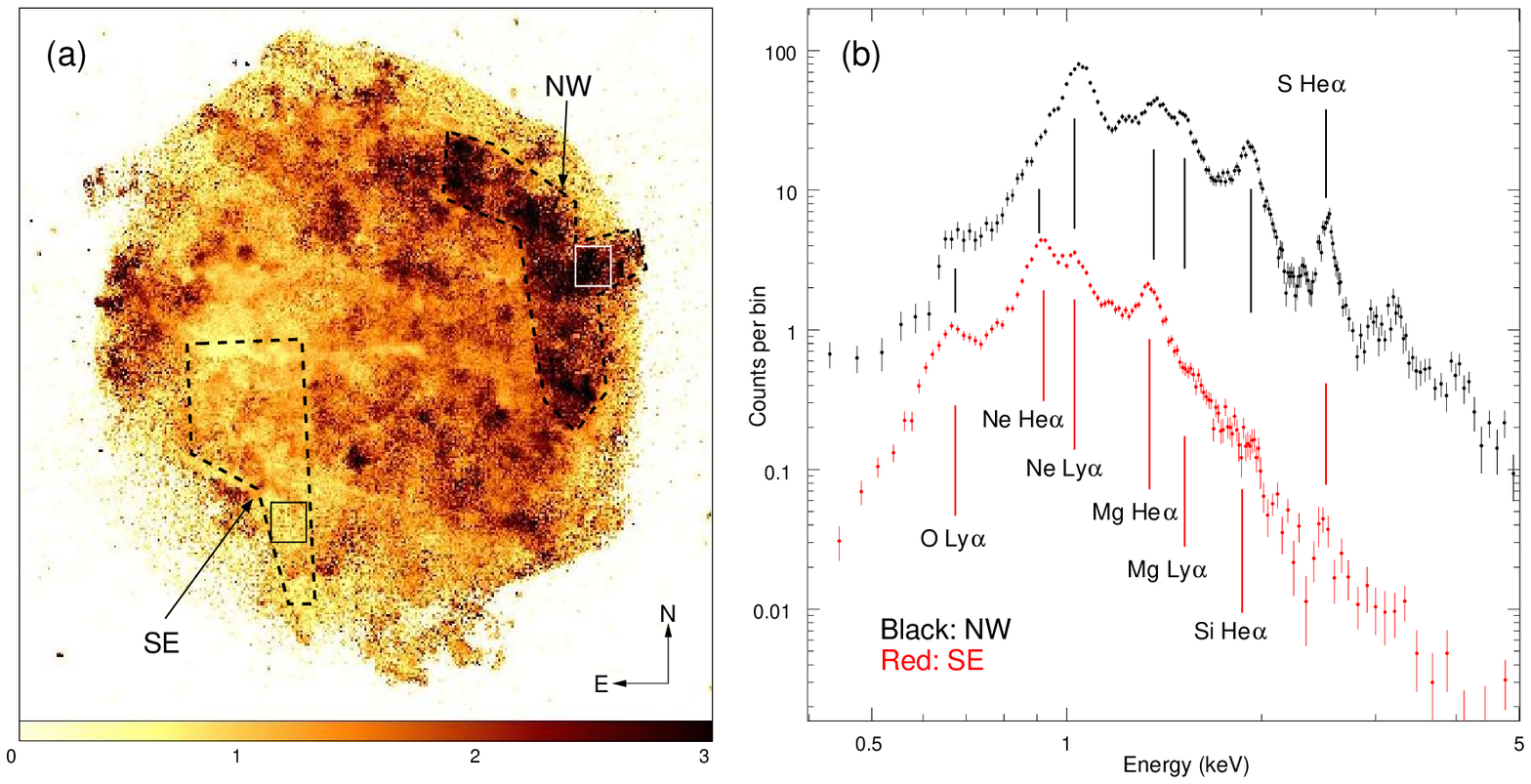}}
\figcaption[]{(a) A hardness radio map of G292.0+1.8: HR = (Ne Ly$\alpha$+Mg+Si+S)/(O+Ne 
He$\alpha$+continuum[$E$$<$1.25 keV]). Energy bands used for line and continuum images 
are presented in Table~\ref{tbl:tab1}. Each subband image has been binned with 
4$\times$4 pixels. (b) Example spectra representing ``NW'' (black) and ``SE'' 
(red) regions. Each spectrum has been arbitrarily scaled for the purposes of display.
In (a), the small region where each spectrum was extracted is marked with a solid box.
Regions outlined by the dashed polygon correspond to the location of bright diffuse 
[O {\small III}] emission (``SE'' region), and the highest HR (``NW'' region).
\label{fig:fig2}}
\end{figure}

\end{document}